# Electrical writing of magnetic and resistive multistates in CoFe films deposited onto Pb[Zr$_x$Ti$_{1-x}$]O$_3$

V. Iurchuk, B. Doudin, J. Bran and B. Kundys

*Institut de Physique et Chimie des Matériaux de Strasbourg (IPCMS), UMR 7504 CNRS-UdS*
*23 rue du Loess, 67034 Strasbourg, France*
*vadym.iurchuk@ipcms.unistra.fr, bdoudin@ipcms.unistra.fr, julien.bran@ipcms.unistra.fr,*
*bohdan.kundys@ipcms.unistra.fr*

**Abstract**
Electric control of magnetic properties is an important challenge for modern magnetism and spintronic development. In particular, an ability to write magnetic state electrically would be highly beneficial. Among other methods, the use of electric field induced deformation of piezoelectric elements is a promising low-energy approach for magnetization control. We investigate the system of piezoelectric substrate Pb[Zr$_x$Ti$_{1-x}$]O$_3$ with CoFe overlayers, extending the known reversible bistable electro-magnetic coupling to surface and multistate operations, adding the initial state reset possibility. Increasing the CoFe thickness improves the magnetoresistive sensitivity, but at the expenses of decreasing the strain-mediated coupling, with optimum magnetic thin film thickness of the order of 100 nm. The simplest resistance strain gauge structure is realized and discussed as a multistate memory cell demonstrating both resistive memory (RRAM) and magnetoresistive memory (MRAM) functionalities in a single structure.

*Keywords:* straintronics, piezoelectricity, ferroelectricity, magnetoelasticity, magnetoresistance, resistive memory

## 1 Introduction

The need of miniaturization of memory bits with the related quest of lowering the power consumption inspire scientific community for searching new ways of controlling magnetic states in spintronic logic devices (Binek and Doudin 2005; Bibes and Barthélémy 2008; Heron et al. 2014). One of such possibility is to use strain as an additional degree of freedom in artificial magnetoelectric structures (Brandlmaier et al. 2011; Brandlmaier et al. 2012; Jia et al. 2012; Roy et al. 2012; Vaz 2012; Liu et al. 2013; Roy 2013; Barangi and Mazumder 2014; Kundys et al. 2014; Li et al. 2014; Iurchuk et al. 2014; Roy 2015). Straintronics (or strain controlled spintronics) is emerging as a new approach, aiming at strain-mediated control of spin dynamics in magnetic media. This approach combines manipulations of ferromagnetic, ferroelectric and ferroelastic orderings in a single working logic element. Importantly, this concept has prominent perspectives for scaling memory bits down to nanoscale range (Rüdiger et al. 2005; Ivry et al. 2009). There is however a need of proof-of-principle experimental insight, showing that such multifunctionality is indeed possible, and ideally providing insights into possible unique device properties of such hybrid materials designs.

Most reported studies in the field of straintronics involve either large or bulk electric fields applied to piezoelectric substrates, in order to obtain two-states magnetization control related to the two electric polarization states of the ferroelectric element. However, multistate non-volatile operation is highly attractive for high-density information storage needs. In that context, our recent reports on the resistive readout of bulk (Kundys et al. 2014) and especially surface (Iurchuk et al. 2014) multilevel strain dynamics of ferroelectric substrate constitute a promising approach for both resistive and magnetoresistive based memories (RRAMs and MRAMs), making possible the combination of both RRAMs functionalities in a single memory cell for multi-bit information storage. Here, a device with the simplest resistance strain gauge design is realized as a multistate memory cell with high reproducibility and reliability for non-volatile operations. The possibility to write, read and erase the remnant states in the device by electric field pulses is demonstrated, while the scientific community is mostly concentrated on switching at constant given voltage. We propose to take advantage of lateral electrodes design (Figure 1) for stressing the piezoelectric substrate, providing the opportunity of lower electric field control. We also detail how electroresistive and magnetoresistive properties of the ferromagnetic element are intertwined. The thickness dependence of resistive and magnetoresistive behavior of ferromagnetic layer is investigated. It has been shown, that reducing the thickness one can obtain better distinguishability of remnant (magnetically coupled) resistive states and increase their number as well. The optimal allowable thickness range is assessed.

## 2 Samples preparation and experimental details

Figure 1 shows the schematic view of the simple planar structure for electro- and magnetoresistive operation. The series of $Co_{0.5}Fe_{0.5}$ films of different thicknesses (20 nm, 70 nm and 150 nm) was grown on the commercially available ceramic substrates of lead zirconium titanate ($Pb(Zr_xTi_{1-x})O_3$ or PZT) (Piezo Ceramic Technology, Piezo Actuators & Piezo Components from PI Ceramic, http://piceramic.com/). CoFe thin films appear to have relatively large magnetostrictive coefficient (67±5 p.p.m. after (Hunter et al. 2011)). The substrates were first mechanically polished to mirror using diamond disk with 3 μm grain size. X-Ray energy dispersion spectroscopy (EDS) under a scanning electron microscope analysis confirmed that PZT does not contain any magnetic impurities within our 0.6% detectability threshold. Thin films were obtained by electron beam evaporation under vacuum in the $10^{-7}$ mbar range, at a rate of 0.1 nm/s. A 10 nm Cr film was used as adhesion layer. The CoFe film was covered by 3 nm of Cr to prevent oxidizing process. We used a shadow mask for patterning the 950×40 μm² CoFe film stripe providing an electroresistive response related to the strain induced geometrical changes of the stripe (du Trémolet de Lacheisserie 1993).

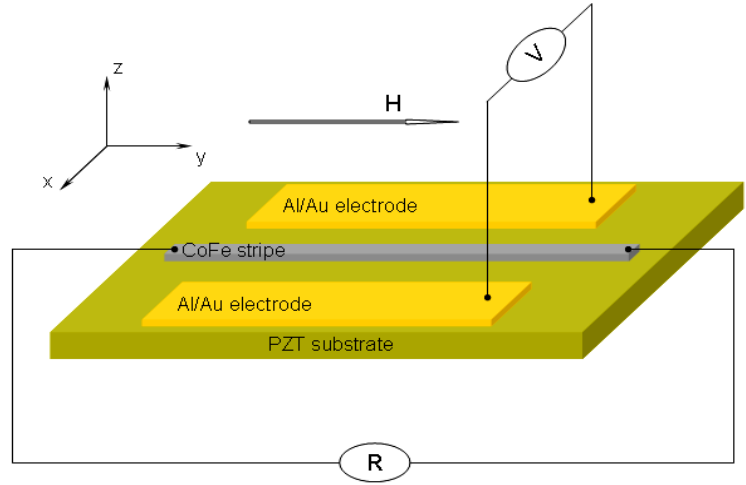

**Figure 1:** The schematics of the experiment. A piezoelectric substrate has a deformation controlled by the stress applied voltage between side Al/Au electrodes, with a CoFe stripe used as a resistive strain gauge as well as a device showing strain-sensitive anisotropic magnetoresistance properties.

Side electrodes made of Al/Au were used to stress the substrate with an electric field. The electrodes were connected using silver paste and the resistance of the film was measured by a 2-probe technique with a LCR meter at the AC voltage of 100mV rms at 100 kHz frequency. All measurements were performed at room temperature.

Magnetoresistive response was measured in the bore of an electromagnet under sweeping field of 3 kOe amplitude, applied at an angle the -120 to 120 degrees range. Figure 2 illustrates the anisotropic magnetoresistive response of a 70 nm CoFe stripe on PZT, which has the expected behavior of a polycrystalline sample.

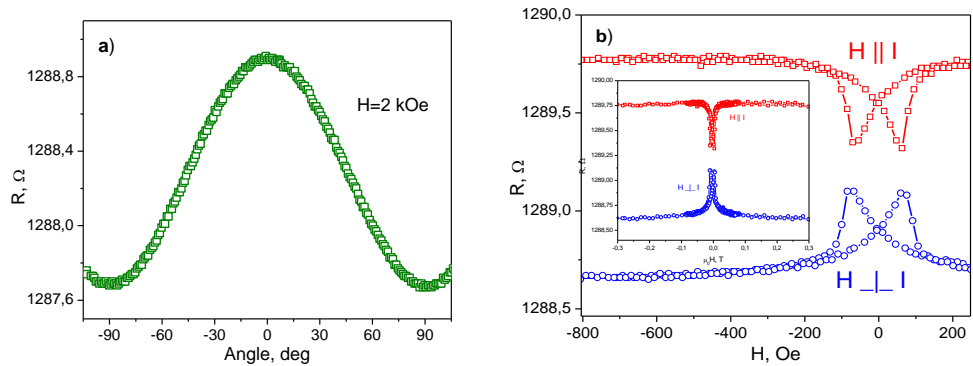

**Figure 2:** Anisotropic magnetoresistance of 70 nm CoFe stripe on PZT versus the angle between H and I (a) and magnetic field amplitude (b).

## 3 Results and discussion

As it is shown in figure 3(a-c), the resistance of the deposited $Co_{0.5}Fe_{0.5}$ stripes exhibit well defined hysteresis loops as a function of the voltage applied between the two side electrodes. The electric field, applied to the sample in an initial $R_0$ state, results in a remnant state different from $R_0$ after returning to zero. By changing the maximum electric field applied ($E_m$) other numerous stable remnant states $R_1,…R_N$ can be successfully created. Notably, the initial state $R_0$ is recoverable with the use of a specific oscillating damped voltage (see below). We have confirmed the stability of these states over time (~$10^6$ s) as well as a good reproducibility of switching processes (~$10^5$ cycles were tested without signs of fatigue).

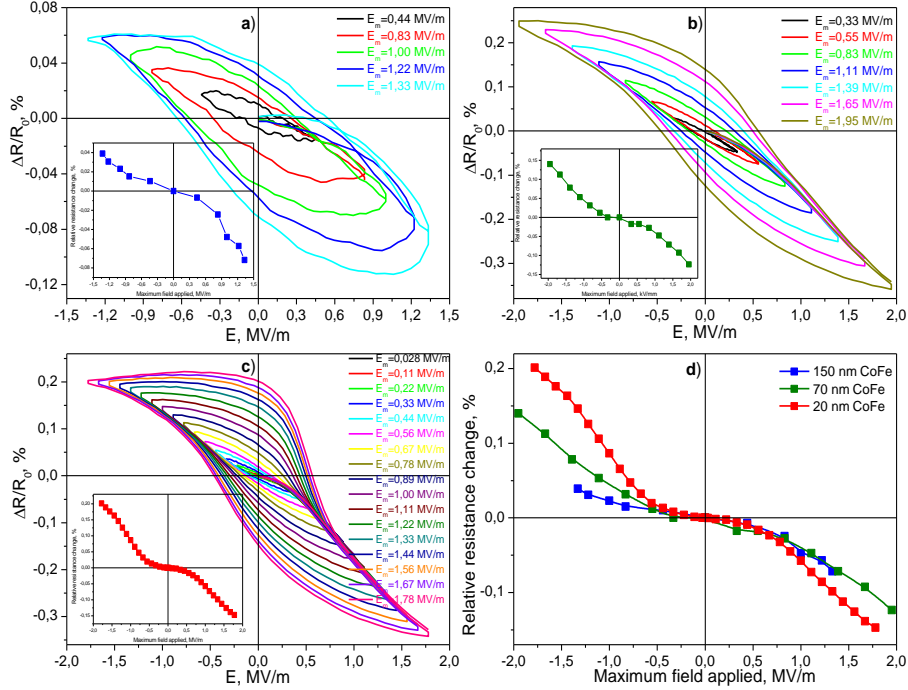

**Figure 3:** Electroresistive hysteresis of CoFe for 150 nm (a), 70 nm (b) and 20 nm (c) thick stripes with multiple memory states at zero electric field for different maximum fields applied to PZT substrate. The insets show relative resistance changes versus maximum field $E_m$. (d) Evolution of the $\Delta R/R_0$ as a function of $E_m$ for different CoFe thicknesses.

The resistive hysteresis originates from the strain hysteresis (figure 4) under small fields cycling. (Kundys et al. 2014) As the PZT compound is near the morphotropic phase boundary, it possesses large converse piezoelectric coefficient, so the deformation remains noticeable even at the small fields applied to PZT. For CoFe stripe tightly bounded to PZT surface, the strain in PZT is directly transferred to CoFe and changes its resistance by shape elasticity means. As the lateral electric field across the stripe ($E_x$) increases, it generates the elastic tensile strain (figure 4), which stretches the stripe transversely and shrinks it along, manifesting herewith the resistance decrease. When $E_x$ sweeps down to zero, the remnant strain configuration in PZT is formed and creates the remnant resistance state, lower than the initial (figure 3(a-c)). Further sweeping to the negative field values generates the compressive strain across the stripe and leads to resistance increase and subsequent creation of another remnant state (higher than the initial), when the field goes back to zero.

The hysteretic R(E) loops are highly reproducible and do not vary for different samples, as long as the geometry of the structure remains the same. However possible slight differences could result from differences in the initial ferroelectric state of the substrate and its ferroelectric history, as it is intrinsically strain-hysteretic.

One has to note that the main contribution to the strain hysteresis is formed from the non-180° ferroelectric domain walls movements, which occur at low electric fields (Damjanovic 2006). Thus working at subcoercive fields region, without contributions of the 180◦ walls displacements, allows avoiding the entire poling of PZT surface, which reduces the ceramics fatigue and is therefore expected to extend the lifetime of the system.

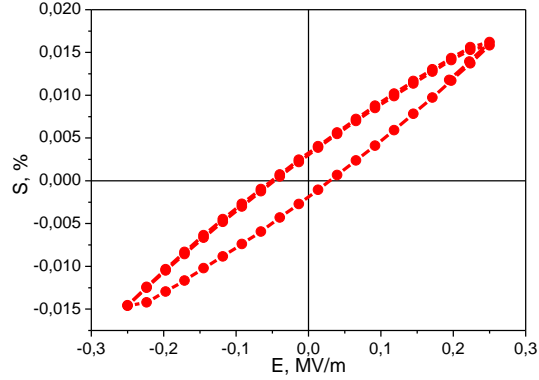

**Figure 4:** Strain vs electric field dynamics in the low electric field regime.

Figure 5 illustrates the diversity and stability of remnant resistive states with an additional possibility of recovering the initial spontaneous elastic strain state in PZT by using a damping oscillating electric field profile applied to a substrate to recreate the close-to-virgin ferroelectric domains distribution. This reset of the initial resistance $R_0$ is extremely important as it allows not only creating the different memory states but also erasing the sample history as well. It appears to be a ferroelectric analogy of the degaussing of ferromagnetic materials with remnant magnetization. As each resistive state is remnant and it corresponds to remnant electric field induced deformation in PZT, the oscillating and dampened electric field restores the close-to-origin surface ferroelectric and, therefore, ferroelastic domains distribution, leading to initial strain (resistive) state reset.

As the technological purposes require ultrahigh speeds for the writing and reading processes, the recent experiments demonstrate, that a deformation response in ferroelectric materials can be of picosecond range (Chen et al. 2012), which would benefit in the ultrafast device performance.

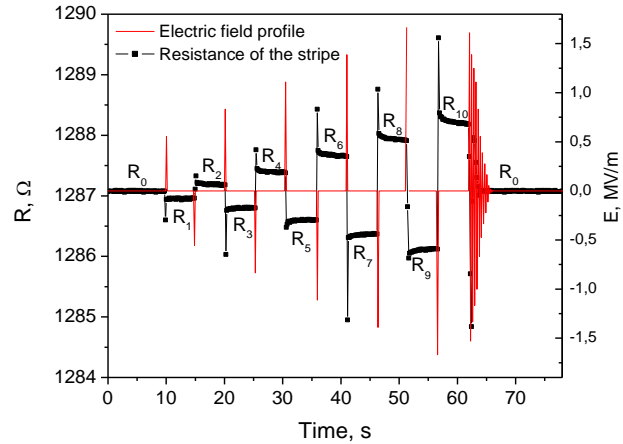

**Figure 5:** Electric field induced resistive switching in PZT/CoFe(70 nm). 11 different states exhibited including the initial $R_0$. Red line shows switching field (with ~100ms pulse duration) with a damping profile to recover an initial $R_0$ state.

In addition to the geometrical strain gauge control of the resistance of CoFe film by an electric field, the magnetoresistance properties ($MR = \dfrac{|R(0) - R(H_{sat})|}{R(H_{sat})}$) can also be modified. More specifically, each state shown in figure 5 is characterized by its own magnetoresistance curve. As

it is indicated in figure 6(a,b), stressing the substrate with voltage pulse (of 100 ms pulse duration) leads to noticeable MR ratio change, which depends on the sign of the applied electric field as well as the direction of the magnetic field. Indeed a positive electric field suppresses the parallel ($H \parallel I$) MR response manifesting in MR ratio decrease (see an upper red loop in figure 6(a)) and increases the perpendicular ($H \perp I$) MR ratio (see a lower red loop in figure 6(a)). The MR dynamics after negative voltage pulse is opposite indicating MR increase for $H \parallel I$ and decrease for $H \perp I$ respectively. We attribute this behavior to the strain-mediated magneto-electric effect that drives the electric field induced magnetic domains redistribution via magnetoelastic mechanism. The electric field applied to PZT modifies its surface by introducing the deformation and surface charge displacement, which shifts the magnetic domains pinning centers in CoFe overlayer, affecting thereby its entire magnetization. Exploring MR curves more in details one can observe that together with MR ratio there is a change in coercivity of CoFe stripe (figure 6(c)), which also exhibits a behavior inverted when changing the signs of the stress. That is, the positive (negative) voltage pulse shrinks (widens) a loop when $H \perp I$ and widens (shrinks) it when $H \parallel I$. The magnetic anisotropy of the CoFe stripe depends on the polarity of the side electric field stress: for positive voltages, a tensile strain is observed (figure 4), which makes the magnetic easy axis more preferentially along the stripe direction. It results in: a) a smaller MR amplitude when the field is parallel to the stripe, signature of an anisotropy increasing along the stripe direction at low magnetic fields, and b) a larger coercive force in the parallel direction, also indicative an increased anisotropy. Concomitantly, the transverse applied magnetic field case shows opposite behaviors.

The magnitude of these effects on resistive and magnetoresistive properties of CoFe overlayer can be improved by varying the CoFe thin film thickness. Thus the thinner stripes demonstrate larger resistance change whereas the MR($\perp$) response is weaker (figure 7), while the change in MR($\parallel$) is negligible. This relates to the fact that thinner films are expected to behave as strain gauge more sensitive to the surface strain as a larger volume fraction of the sample is deformed under strain. The MR properties, however, do not follow the same trend. The evolution of MR with thickness indicates that the anisotropy becomes increasingly stress dependent when increasing the material thickness. One tentative explanation is that the thin film anisotropy decreases with thickness, and it is in the perpendicular direction where these effects are most stress-sensitive. Fine-tuning the thin film fabrication conditions might lead to better control of the anisotropy, possibly optimizing the thickness sensitivity. But this is beyond the scope of our proof-of-principle presentation here. The MR ratio thickness dependence of opposite trend sets an optimal thickness range for CoFe stripes on PZT to 100±20 nm.

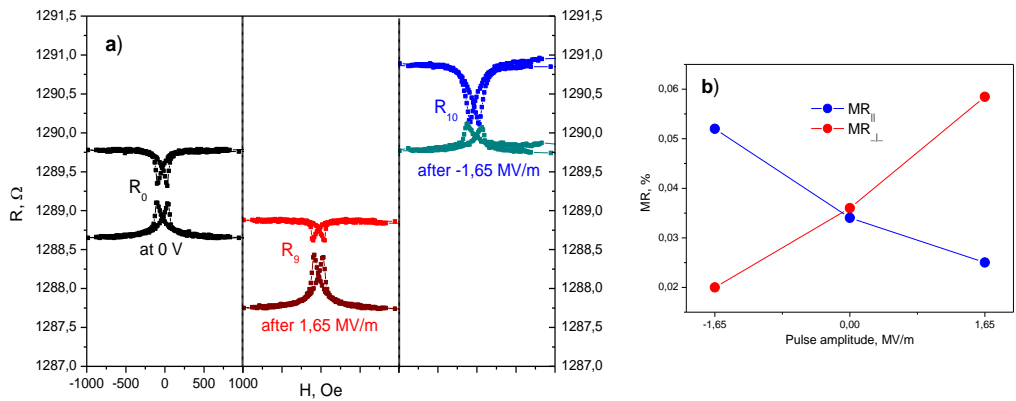

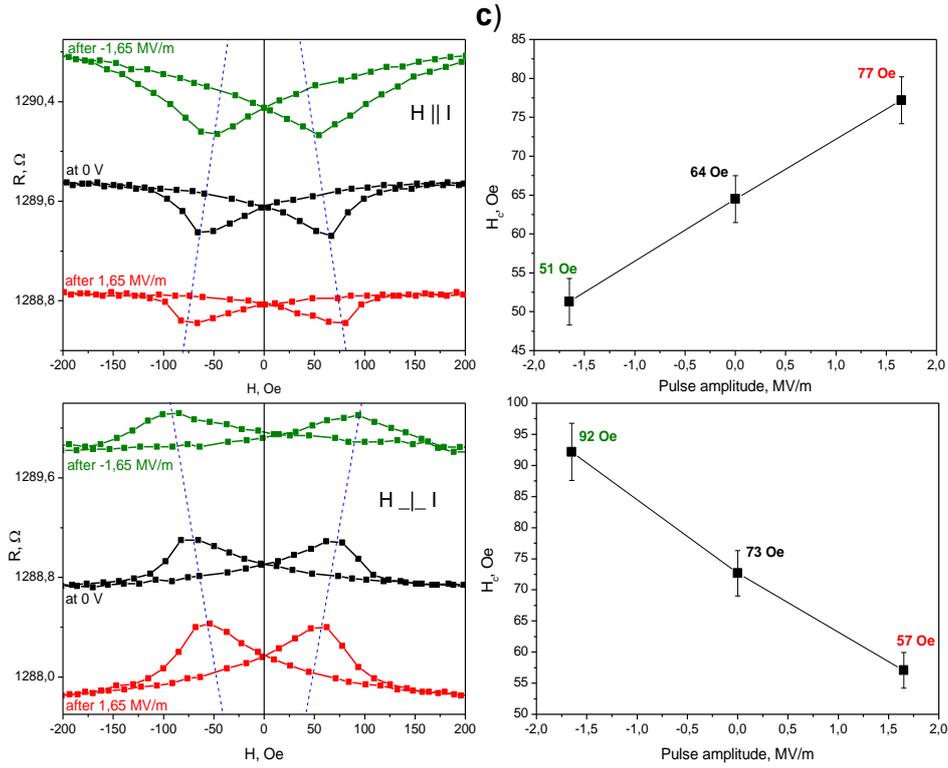

**Figure 6:** 70 nm CoFe stripe magnetoresistance after electric field pulse applied to PZT indicating MR ratio change (a, b) as well as coercive field modification (c).

It has to be noticed that the values of ΔR/R and MR can be considerably improved using a ferroelectric single crystals or epitaxially grown thin ferroelectric films with larger piezoelectric coefficient and high-quality surface to provide better strain transfer to ferromagnetic thin film. Moreover there is a possibility to use different geometrical configurations of the strain gauge (i.e. zigzag or spiral) to increase the gauge factor. Although the voltages needed for switching are still relatively high, they can be sufficiently lowered with scaling down the structures. Using this design it is possible to scale down the structure to submicron range. Significant size reducing would lead to lowering the voltages needed for switching and, probably, would increase the magnitude of observed remnant effects. One should expect that sub-100nm ferroelastic domain structures can be stabilized and controlled in micron (Paruch et al. 2001; Balke et al. 2009) or even submicron cells (Yamada et al. 2013; Boyn et al. 2014), which will not only increase the resolution between different states, but will also allow using operating voltages less than 1 V.

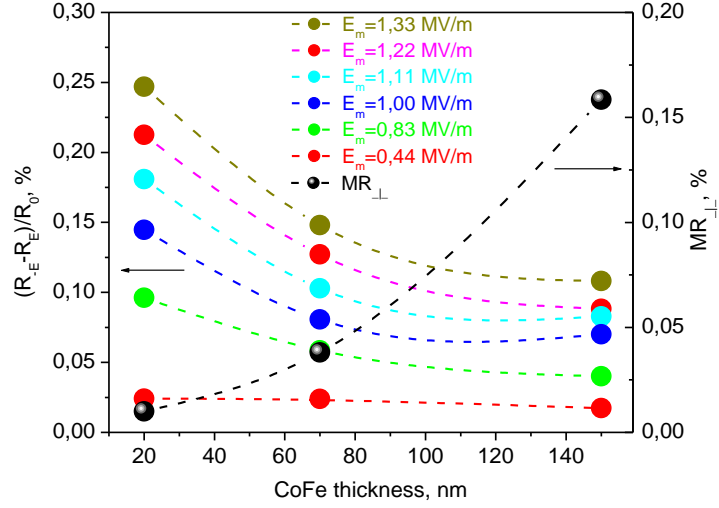

**Figure 7:** $\delta R=(R(-E_m)-R(E_m))/R_0$ ratios for different $E_m$ excitations and MR($H \perp I$) ratio as the function of CoFe stripe thickness. Dash lines are guides to the eye.

Although the values of the reading current are still in the microampere range, much smaller currents are expected when miniaturizing, and by using the higher impedance (TMR) reading cells, which is beyond the scope of our proof-of-principle experiment. With GMR or TMR reading, the magnetic orientation discrimination requires spin valves engineering, with the bottom layer having proper anisotropy tunable by substrate strain. Such device design testing is currently in progress.

# 4 Conclusions

We showed that the deterministic control of ferroelectric surface strain that allows ferroelastic writing of magnetic and resistive multistates at room temperature. Combining piezoresistive and magnetoelastic effects that couple substrate deformation with charge and magnetic transport characteristics makes possible the realization of hybrid straintronic-spintronic structures exhibiting both RRAM and MRAM functionalities in single sample when the strain gauge is made of a magnetostrictive CoFe material. The optimal thickness of the FM film of the prototype device has been determined to be 100±20 nm. Further miniaturization and optimization of this simplified device concept by implementation of different materials (ferroelectric and multiferroic monocrystalline or polymer substrates together with the spin valves as the resistive sensors) would enhance the low power spintronic devices and their permanent storage capabilities.

# Acknowledgments

We gratefully acknowledge the research grants from Agence Nationale de la Recherche (hvSTRICTSPIN ANR-13-JS 04-0008-01, Labex NIE 11-LABX-0058-NIE, Investissement d'Avenir programme ANR-10-IDEX-0002-02) for financial support. We are immensely grateful to Mr J. Faerber who provided analysis of the PZT substrates. The technical help of Mr H. Majjad (STnano cleanroom), Mr B. Leconte and Mr F. Chevrier is also gratefully acknowledged.